\documentclass[12pt]{iopart}
\usepackage{iopams} 
\usepackage[usenames,dvipsnames]{xcolor} 
\usepackage[normalem]{ulem}
\usepackage{graphicx}

\begin{document}

\title[String analog of Reissner-Nordstr\"{o}m black holes cannot be overcharged]{String analog of Reissner-Nordstr\"{o}m black holes cannot be overcharged}

\author{Koray D\"{u}zta\c{s}}

\address{Physics Department, Eastern Mediterranean  University, Famagusta, North Cyprus via Mersin 10, Turkey}
\address{Department of Natural and Mathematical Sciences\\
\"{O}zye\u{g}in University 34794 \.{I}stanbul Turkey
}
\ead{koray.duztas@ozyegin.edu.tr}
\author{Mubasher Jamil}
\address{Department of Mathematics, School of Natural Sciences (SNS), National
University of Sciences and Technology (NUST), H-12, Islamabad, Pakistan}
\address{Institute of Astrophysics \\ Zheijiang University of Technology, Hangzhou China}
\ead{mjamil@zjut.edu.cn}
\vspace{10pt}
%\begin{indented}
%\item[]February 2014
%\end{indented}

\begin{abstract}
In this work we attempt to overcharge extremal and nearly extremal charged black holes in string theory, known as the Garfinkle-Horowitz-Strominger solution.  We first show that extremal black holes cannot be overcharged analogous to the case of Reissner-Nordstr\"{o}m black holes. Contrary to their analogues in general relativity, nearly extremal black holes can neither be overcharged beyond extremality, nor can they be driven to extremality by the interaction with test particles. Therefore the analysis in this work also imply that the third law of black hole thermodynamics holds for the relevant charged black holes in string theory perturbed by test particles. This can be interpreted as a stronger version of the third law since one can drop out the continuity proviso for the relevant process.
\end{abstract}

% Uncomment for PACS numbers
%
% Uncomment for keywords
\vspace{2pc}
\noindent{\it Keywords}: Black holes in string theory; Cosmic censorship; black hole thermodynamics
%
% Uncomment for Submitted to journal title message
%\submitto{\CQG}
%
% Uncomment if a separate title page is required
%\maketitle
% 
% For two-column output uncomment the next line and choose [10pt] rather than [12pt] in the \documentclass declaration
%\ioptwocol
%
\section{Introduction}
Static, spherically symmetric and charged black holes in general relativity are identified by their mass, charge and angular momentum parameters. The singularities at the center of these black holes are covered by event horizons provided that the parameters satisfy the relevant inequalities. However as the curvature approaches the Planck scale, general relativity is expected to break down while the quantum effects start to dominate. String theory is the most promising candidate to incorporate quantum effects into gravity, which would dominate near the singularities. The fact that the low energy string theory predicts the existence of static and charged black holes lends credence to its validity as a quantum theory of gravity. 

The Schwarzschild solution in general relativity is a good approximation to static, uncharged black holes in string theory; especially outside the event horizon where the curvature is sufficiently low. The predictions of string theory do not considerably deviate from Schwarzschild solution until one approaches the singularity at the center of the black hole. However, the fact that the dilaton in heterotic string theory couples to the Maxwell field precludes charged black hole solutions with a constant dilaton. Therefore the Reissner-Nordstr\"{o}m (RN) solution in general relativity does not overlap with charged black holes in low energy string theory, even at the asymptotically flat limit. 

The action for the fields in the low energy limit of heterotic string theory is
\begin{equation}
S=\int d^Dx \sqrt{-g} e^{-2\phi} [ \Lambda + R + 4(\nabla \phi)^2 - F^2-(1/2) H_{\mu \nu \rho}H^{\mu \nu \rho}] \label{action1}
\end{equation}
where $\phi$ is the dilaton, $F_{\mu \nu}$ is the Maxwell field, and the three form $H_{\mu \nu \rho}$ is defined by $dH=-F\wedge F$. $\lambda$ is a constant which depends on the spacetime dimension.  We intend to simplify the action (\ref{action1}) and derive black hole solutions. For that purpose we restrict ourselves to four spacetime dimensions, and impose the boundary conditions that the spacetime is asymptotically flat and the dilaton approaches  zero at infinity. Setting $\Lambda=0$, and $H=0$ yields 
\begin{equation}
S=\int d^4 x \sqrt{-g^E}(R_E -2 (\nabla \phi)^2- e^{-2\phi}F^2 ) \label{action2}
\end{equation}
Note that the metric in (\ref{action1}) is rescaled by $e^{-2\phi}$, i.e. $g^E_{\mu \nu}=e^{-2\phi}g_{\mu \nu}$, to ensure that   (\ref{action2}) reduces to the standard Einstein action with scalar field, when $F_{\mu \nu}=0$. This theory involves remarkably simple solution black hole solutions which can be derived either by solving the field equations or employing solution generating techniques (see e.g.~\cite{horowitz})
\begin{eqnarray}
&&ds^2_E=-\left( 1- \frac{2M}{r} \right) dt^2 + \left(  1- \frac{2M}{r} \right)^{-1} dr^2 +r\left( r-\frac{Q^2}{M} \right)d\Omega \nonumber \\
&& F_{rt}=\frac{Q}{r^2}; \quad e^{2\phi}=1-\frac{Q^2}{Mr}
\label{stringrn}
\end{eqnarray}
The solution (\ref{stringrn}) which describes a charged black hole in string theory (CBHST), was first derived by Gibbons~\cite{gibbons1}, who also elaborated on it with Maeda~\cite{gibbons2}. Later, it was independently found by Garfinkle, Horowitz, and Strominger~\cite{garfinkle}. The spatial coordinate of the event horizon of this black hole is fixed at $r=2M$. There exists a singular surface at $r=Q^2/M$, which is covered by the event horizon provided that $Q^2<2M^2$. The spatial coordinate of the event horizon does not depend on the charge as in the case of RN black holes. However, as we increase the charge $Q$, the singular surface moves out in the spatial coordinate $r$, and coincides with the event horizon in the extremal limit $Q^2=2M^2$. If one can increase $Q$ further, the singularity moves outside the horizon and becomes timelike. 

In this work we investigate if it is possible to increase the charge $Q$ of a CBHST beyond the extremal limit, in the spirit of cosmic censorship conjecture proposed by Penrose~\cite{ccc}. This conjecture asserts that the singularities which inevitably form in gravitational collapse should be covered by event horizons which disable their causal contacts with the spacetime outside the black hole region. This way the smooth structure of the spacetime is maintained at least in the region outside the event horizon. In the absence of a concrete proof, an alternative method to test  the stability of event horizons was proposed by Wald. In Wald type problems, one starts with an extremal or a nearly extremal black hole and evaluates the possibility to overcharge or overspin the black hole into a naked singularity. In the first of these thought experiments Wald showed that particles with sufficiently large charge or angular momentum to destroy an extremal Kerr-Newman black hole are not absorbed by the black hole~\cite{wald74}. Hubeny adopted a more subtle approach where one starts with a nearly extremal black hole and showed that RN black holes can be overcharged in this manner~\cite{hu}. The stability of event horizons was tested for various cases in Einstein-Maxwell theory ~\cite{f1,gao,magne,dilat,higher,v1,w2,mu}. It is also possible to test the validity of the cosmic censorship conjecture in the case of test fields scattering off black holes~\cite{dkn,overspin,emccc,duztas,toth,natario,hawk,mode,btz,taub-nut,gwak3}. Recently, similar thought experiments were constructed involving the perturbations of Kerr-Sen black holes in heterotic string theory by test particles and fields~\cite{siahaan,kerrsen}. Here, we construct thought experiments by sending charged particles to extremal and nearly extremal CBHST's and evaluate whether they can be overcharged beyond extremality; i.e. whether the singularity can become naked. Our analysis also allows us to test the validity of the third law of black hole dynamics, which states that a nearly extremal black hole cannot become extremal in any continuous process.

\section{Overcharging string analogue of R-N black holes}
In thought experiments to overcharge or overspin black holes we send in test particles or fields to the black hole from infinity. We assume that the interaction with the test particle does not change the geometry of the spacetime fundamentally, but leads to perturbations in mass, charge and angular momentum parameters. We check whether the final parameters of the spacetime  represent a black hole or  a naked singularity. For that purpose we should first demand that the test particle is absorbed by the black hole. This condition entails the existence of a lower limit for the energy of the test particle. 
To find the lower limit, let us  consider the equations of motion of a test particle of mass $m$ charge $q$  
\begin{equation}
\ddot{x}^{\mu}+ \Gamma^{\mu}_{\rho \sigma}\dot{x}^{\rho} \dot{x}^{\sigma}=\frac{q}{m}F^{\mu \nu}\dot{x}_{\nu}
\end{equation}
which can be derived from the Lagrangian
\begin{equation}
\mathcal{L}=\frac{1}{2}mg_{\mu \nu}\dot{x}^{\mu}\dot{x}^{\nu}+ qA_{\mu}\dot{x}^{\mu}
\end{equation}
where $A=(-Q/r) dt$ for the solution (\ref{stringrn}), and $q$ is the charge of the test particle. The energy of a particle is a conserved quantity which is given by
\begin{equation}
E=-\frac{\partial \mathcal{L}}{\partial \dot{t}}=-mg_{00}\dot{t}-qA_0=m\left(1-\frac{2M}{r}\right)\dot{t}+\frac{qQ}{r} \label{E}
\end{equation}
 The equation (\ref{E}) implies that at $r=r_+=2M$, the minimum energy is 
\begin{equation}
E\geq E_{\rm{min}}= \frac{qQ}{r_+}=\frac{qQ}{2M} \label{emin}
\end{equation}
The particles with energy less than $E_{\rm{min}}$ will never cross the horizon to be absorbed by the black hole. We proceed by evaluating whether a test particle with $E>E_{\rm{min}}$ can overcharge a CBHST beyond the extremal limit $Q^2=2M^2$. We first evaluate the extremal case.

\subsection{Extremal case}
We attempt to overcharge an extremal black hole by sending in test particles with charge $q=Q\epsilon$, where $\epsilon \ll 1$. Initially the extremal black hole satisfies
\begin{equation}
\delta_{\rm{in}}=2M^2-Q^2=0 \label{deltain}
\end{equation}
First, to be absorbed by the extremal black hole, a test particle should satisfy
\begin{equation}
E\geq E_{\rm{min}}=\frac{qQ}{2M}=\frac{\epsilon Q^2}{2M}=M\epsilon \label{condi1ex}
\end{equation}
where we have substituted $Q^2=2M^2$, and $q=Q\epsilon$, in the expression for $E_{\rm{min}}$.
To overcharge the black hole we also demand that 
\begin{equation}
\delta_{\rm{fin}}=2(M+E)^2-(Q+q)^2<0 \label{deltafin}
\end{equation}
where $E$ is the energy of the test particle and $q=Q\epsilon$ is its charge. Note that the energy and the charge of the test particle contribute to the mass and the charge parameters of the black hole, respectively. Let us re-write (\ref{deltafin}) by substituting $Q^2=2M^2$, and $q=Q\epsilon$.
\begin{equation}
2E^2+4ME-2M^2(\epsilon^2 + 2\epsilon)<0
\end{equation}
$\delta_{\rm{fin}}$ reduces to a quadratic equation for $E$, which has two roots $E_1,E_2$. $\delta_{\rm{fin}}$ will be negative for $E_1<E<E_2$ and positive outside the range $(E_1,E_2)$. The roots are given by
\begin{eqnarray}
E_{1,2}&=& \frac{-4M \pm \sqrt{16M^2+4.2.(2M^2)(\epsilon^2 + 2\epsilon)}}{4} \nonumber \\
&=& -M \pm M(1+\epsilon)\label{rootex1}
\end{eqnarray}
$\delta_{\rm{fin}}$ has a positive and a negative root. We demand that $\delta_{\rm{fin}}$ is negative so that the final parameters of the black hole represent a naked singularity. This will be possible if the energy of the test particle is chosen in the range $(E_1,E_2)$. Therefore the maximum energy for the test particle (with charge $q=Q\epsilon$) to overcharge the extremal black hole is
\begin{equation}
E<E_{\rm{max}}=E_2=M\epsilon \label{condi2ex}
\end{equation}
The conditions (\ref{condi1ex}) and (\ref{condi2ex}) cannot be satisfied simultaneously since $E_{\rm{min}}=E_{\rm{max}}$. The test particles with energy $E<E_{\rm{max}}$ which could overcharge the extremal black hole, are not absorbed by the black hole. Apparently it is not possible to overcharge an extremal CBHST. This is analogous to the case of RN black holes.

\subsection{Nearly Extremal Black Holes}
In this section we attempt to overcharge a nearly extremal CBHST, to compare the results with nearly extremal RN black holes which can be overcharged if backreaction effects are neglected.
We parametrise a nearly extremal black hole as
\begin{equation}
\delta_{\rm{in}}=2M^2-Q^2=M^2\epsilon^2 \label{param}
\end{equation}
The minimum energy for a test particle with charge $q=Q\epsilon$ to be absorbed by this nearly extremal black hole is given by
\begin{equation}
E\geq E_{\rm{min}}=\frac{qQ}{2M}=\frac{\epsilon Q^2}{2M}=M\epsilon \left(1-\frac{\epsilon^2}{2} \right) \label{condi1nex}
\end{equation}
where we have substituted $Q^2=M^2(2-\epsilon^2)$, and $q=Q\epsilon$, in the expression for $E_{\rm{min}}$. Again we demand that 
$\delta_{\rm{in}}<0$ at the end of the interaction so that the nearly extremal black hole is overcharged. 
\begin{equation}
\delta_{\rm{fin}}=2(M+E)^2-(Q+q)^2<0 \label{deltafinnex}
\end{equation}
Substituting $Q^2=M^2(2-\epsilon^2)$, and $q=Q\epsilon$ in (\ref{deltafinnex}), we get
\begin{equation}
2E^2+ 4ME -M^2(\epsilon^2-\epsilon^4 +4\epsilon -2\epsilon^3)<0
\end{equation}
Again $\delta_{\rm{fin}}$ reduces to a quadratic equation for $E$, which has two roots $E_1,E_2$. $\delta_{\rm{fin}}$ will be negative for $E_1<E<E_2$ and positive outside the range $(E_1,E_2)$. The roots are given by
\begin{eqnarray}
E_{1,2}&=& \frac{-4M \pm \sqrt{16M^2+8M^2(\epsilon^2-\epsilon^4 +4\epsilon -2\epsilon^3)}}{4} \nonumber \\
&=& M \left(-1\pm \sqrt{1+2\epsilon + \frac{\epsilon^2}{2}-\epsilon^3 -\frac{\epsilon^4}{2}}\right)\label{rootnex1}
\end{eqnarray}
If the energy of the test particle is chosen in the range $(E_1,E_2)$, $\delta_{\rm{fin}}$ will be negative which means that the nearly extremal black hole is overcharged. The maximum energy for the test particle (with charge $q=Q\epsilon$) to overcharge the nearly extremal black hole is
\begin{equation}
E<E_{\rm{max}}=E_2=M\left(\sqrt{1+2\epsilon + \frac{\epsilon^2}{2}-\epsilon^3 -\frac{\epsilon^4}{2}}-1\right)\label{condi2nex}
\end{equation}
The test particle should satisfy the two conditions (\ref{condi1nex}) and (\ref{condi2nex}) simultaneously so that it is absorbed by the nearly extremal CBHST to overcharge it. This could be possible if $E_{\rm{max}}$ given in (\ref{condi1nex}) is larger than $E_{\rm{min}}$ given in (\ref{condi2nex}). However the maximum energy for the nearly extremal case is less than the minimum energy to be captured by the black hole. The particles that could overcharge the black hole are not absorbed by the black hole. To see this explicitly, notice that the condition $E_{\rm{max}}>E_{\rm{min}}$ is equivalent to
\begin{equation}
1+2\epsilon + \frac{\epsilon^2}{2}-\epsilon^3 -\frac{\epsilon^4}{2}>\left(\epsilon-\frac{\epsilon^3}{2}+1 \right)^2
\end{equation}
which implies
\begin{equation}
-\frac{\epsilon^2}{2}+\frac{\epsilon^4}{2}-\frac{\epsilon^6}{4}>0 \label{condinex3}
\end{equation}
which cannot be satisfied by any choice of $\epsilon$ that is real and positive. $E_{\rm{max}}<E_{\rm{min}}$ independent of the choice of $\epsilon$, therefore one cannot find an energy for the test particle to overcharge the nearly extremal CBHST. 
\begin{center}
\begin{figure}
\includegraphics[scale=0.8]{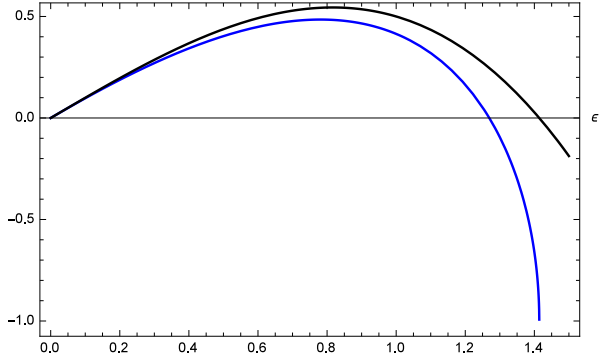}
\caption{The graphs of $E_{\rm{min}}$ and $E_{\rm{max}}$ for $M=1$. Here $\epsilon$ is allowed to vary beyond the test particle limit to clearly visualise that $E_{\rm{min}}>E_{\rm{max}}$}
\end{figure}
\end{center}
Another way to check if the nearly extremal black hole can be overcharged is to substitute $E_{\rm{min}}$ into the expression for $\delta_{\rm{fin}}$ to check if it can be negative.
\begin{equation}
\delta_{\rm{fin}}=2(M+E_{\rm{min}})^2-(Q+q)^2
\end{equation}
Substituting $Q^2=M^2(2-\epsilon^2)$, $q=Q\epsilon$, and $E_{\rm{min}}=M\epsilon(1-\epsilon^2/2)$ in the expression for $\delta_{\rm{fin}}$, we get
\begin{eqnarray}
\delta_{\rm{fin}}&=&(2M^2-Q^2)+2E_{\rm{min}}^2+ 4ME_{\rm{min}}-\epsilon^2 Q^2-2\epsilon Q^2 \nonumber \\
&=& M^2\epsilon^2 -M^2\epsilon^4 +\frac{M^2\epsilon^6}{2} \label{deltafinnex2}
\end{eqnarray}
Manifestly $\delta_{\rm{fin}}$ is positive for the minimum energy which ensures that the test particle is absorbed by the black hole. For larger energies,  $\delta_{\rm{fin}}$ will be larger than the value derived in (\ref{deltafinnex2}). Therefore it will not be possible for any test particle that is captured by the nearly extremal CBHST to overcharge it.

For a numerical example let us consider a nearly extremal CBHST with $M=1$ and choose $\epsilon=0.01$. The parametrization (\ref{param}) implies $Q^2=1.9999$. For this black hole 
\begin{eqnarray}
E_{\rm{min}}&=&M\epsilon \left(1-\frac{\epsilon^2}{2}\right)=0.01 \left(1-\frac{0.01^2}{2}\right) \nonumber \\
&=&0.0099995
\end{eqnarray}
We can calculate $\delta_{\rm{fin}}$
\begin{eqnarray}
\delta_{\rm{fin}}&=&2(M+E_{\rm{min}})^2-(Q+q)^2=2(M+E_{\rm{min}})^2-Q^2(1+\epsilon)^2 \nonumber \\
&=&2(1.0099995)^2-1.9999(1.01)^2 \nonumber \\
&=&0.00009999
\end{eqnarray}
For any particle captured by the black hole the value of $\delta_{\rm{fin}}$ will be at least $\sim 0.0001$. Therefore  test particles cannot overcharge the black hole. The maximum energy for a test particle to overcharge the black hole can also be calculated using (\ref{condi2nex})
\begin{eqnarray}
E_{\rm{max}}&=&M\left(\sqrt{1+2\epsilon + \frac{\epsilon^2}{2}-\epsilon^3 -\frac{\epsilon^4}{2}}-1\right) \nonumber \\
&=& \sqrt{1+2(0.01)+\frac{(0.01)^2}{2}-(0.01)^3-\frac{(0.01)^4}{2}}-1 \nonumber \\
&=&0.00997475
\end{eqnarray}
We see that $E_{\rm{max}}<E_{\rm{min}}$ in accord with the derivation in this section. The test particles with energy below $\sim 0.00997475$ could overcharge the nearly extremal black hole, but they are not absorbed by the black hole since $E<E_{\rm{min}}$, which is an alternative justification of the fact that a nearly extremal CBHST cannot be overcharged, contrary to RN black holes in general relativity.

\section{Validity of the third law of black hole dynamics}
The laws of black hole dynamics were first suggested by Bardeen, Carter, and Hawking for stationary and axisymmetric black holes in general relativity~\cite{bardeen}. The authors discovered a close analogy with the area of the event horizon and the surface gravity $\kappa$ of the black hole with entropy and temperature, respectively. The second law which states that the area of the event horizon cannot be decreased had already been proved by Hawking by assuming the validity of the cosmic censorship conjecture~\cite{area}. The third law which states that $\kappa$ cannot be reduced to zero by a finite sequence of operations, lacked a rigorous proof. Later, Israel gave a formal statement of the third law and a proof  based on his gravitational confinement theorem~\cite{israel0,israel1}. According to the formal statement of the third law, a non-extremal black hole cannot become extremal at a finite advanced time in any continuous process. The proof of the third law pre-assumes that the trapped surfaces can be extended without encountering a singularity, and the extension is semirigid, i.e. the areas of the trapped surfaces are preserved. These assumptions could not be justified in a case  where the cosmic censorship conjecture is violated (see \cite{thermo}). 

In this work we derived that the minimum energy $E_{\rm{min}}$ that allows the absorption of a test body by a nearly extremal CBHST is always larger than (never equal to) the maximum energy $E_{\rm{min}}$ that could lead to the overcharging of the CBHST. In other words $\delta_{\rm{fin}}$ is always greater than (never equal to) zero at the end of the interaction of a nearly extremal CBHST with test particles. Therefore the interactions of non-extremal CBHST's with test particles cannot drive them to extremality. The analysis for the nearly extremal case constitutes proofs both for the validity of the cosmic censorship conjecture and the third law of black hole dynamics for string analogues of R-N black holes interacting with test particles. 

Here, we assumed that the absorption of the energy and charge of test particles occurs at a single step, rather than a continuous process. This could lead to a discrete jump from a nearly extremal black hole to a naked singularity as in the case of RN black holes studied by Hubeny. However, this is not the case for the string analogues of RN black holes. We derived that a nearly extremal CBHST cannot be driven to extremality or beyond, even if it absorbs the energy and charge of test particles discontinuously. From that point of view, a stronger version of the third law holds for the CBHST's interacting with test particles, where one can drop out the continuity proviso for the relevant process.
\section{Summary and conclusions}
In this work we perturbed extremal and nearly extremal CBHST's with test particles to evaluate whether they can be overcharged beyond the extremal limit $Q^2=2M^2$. We derived the minimum energy for a test particle to ensure that it is absorbed by the black hole, and the maximum energy of a test particle that could overcharge the black hole. The particles with charge $q=Q\epsilon$ and energy in the range $(E_{\rm{min}},E_{\rm{max}})$ could be absorbed by the black hole to overcharge it, provided that $E_{\rm{min}}<E_{\rm{max}}$. However, it turns out that $E_{\rm{min}}=E_{\rm{max}}$ for extremal black holes. Therefore extremal CBHST's cannot be overcharged analogous to the case of RN black holes. We also analysed the nearly extremal case to derive that  $E_{\rm{min}}<E_{\rm{max}}$. Nearly extremal CBHST's can neither be driven to extremality nor beyond extremality, contrary to the case of RN black holes.

The fact that the nearly extremal CBHST's cannot be driven to extremality implies that the third law of black hole dynamics holds for CBHST's interacting with test particles. We argued that a stronger version of the third law is implied where one can drop out the continuity proviso for the relevant process. We note that the analysis in this work cannot be interpreted as a general proof for the third law of black hole dynamics for CBHST's, since only a special case is evaluated here.

\end{document}